\documentclass[useAMS,usenatbib]{mn2e}
\usepackage{times}
\usepackage{epsfig}
\usepackage{colortbl,color}

\usepackage[dvipdfm,colorlinks=true,linkcolor=blue,urlcolor=blue,citecolor=blue]{hyperref}

\newcommand{\msun}{{\rm M}_{\sun}}
\newcommand{\ledd}{L_{{\rm Edd}}}

\newcommand{\lx}{L_{\rm X}}

\newcommand{\lbol}{L_{\rm bol}}

\title[$R_{\rm UV}$ -- $\alpha_{\rm ox}$ correlation in LLAGNs]{A strong negative correlation between radio loudness $R_{\rm UV}$ and optical-to-X-ray spectral index $\alpha_{\rm ox}$ in low-luminosity AGNs}
\author[S.-L. Li and F.-G. Xie]{Shuang-Liang Li\thanks{E-mails:lisl@shao.ac.cn (S.L.L.); fgxie@shao.ac.cn (F.G.X.)} and Fu-Guo Xie$^\star$\\
Key Laboratory for Research in Galaxies and Cosmology, Shanghai Astronomical Observatory, Chinese Academy of Sciences,\\
80 Nandan Road, 200030 Shanghai, China}

\begin{document}

\pagerange{\pageref{firstpage}--\pageref{lastpage}} \pubyear{2017}

\maketitle


\begin{abstract}
It has been argued for years that the accretion mode changes from bright active galactic nuclei (AGNs) to low-luminosity AGNs (LLAGNs) at a rough dividing point of bolometric Eddington ratio $\lambda \sim 10^{-2}$. In this work, we strengthen this scenario through investigation of the relationship between the radio loudness $R_{\rm UV}$ and the optical-to-X-ray spectral index $\alpha_{\rm ox}$ in LLAGNs with $10^{-6}\la \lambda \la 10^{-3}$. We compile from literature a sample of 32 LLAGNs, consisting 18 LINERs and 14 low Eddington ratio Seyfert galaxies, and observe a strong negative $R_{\rm UV}$--$\alpha_{\rm ox}$ relationship, with large scatter in both $R_{\rm UV}$ and $\alpha_{\rm ox}$. We further demonstrate that this negative correlation, and the additional two negative relationships reported in literature ($R_{\rm UV}$--$\lambda$ and $\alpha_{\rm ox}$--$\lambda$ correlations), can be understood consistently and comprehensively under the truncated accretion--jet model, the model that has been applied successfully applied to LLAGNs. We argue that the scatter in the observations are (mainly) due to the spread in the viscosity parameter $\alpha$ of a hot accretion flow, a parameter that potentially can serve as a diagnose of the strength and/or configuration of magnetic fields in accretion flows.
\end{abstract}


\section{INTRODUCTION}

It has been known for years that statistically low-luminosity active galactic nuclei (AGNs), whose  bolometric Eddington ratios (defined as the bolometric luminosity $\lbol$ in Eddington unit, i.e. $\lambda\equiv\lbol/\ledd$, where $\ledd \equiv 1.3\times 10^{46} {\rm erg\ s^{-1}} (M_{\rm BH}/10^8 \msun)$ is the Eddington luminosity and $M_{\rm BH}$ is the black hole mass) are lower than $\sim 1-2\times10^{-2}$, are distinctive in various aspects to those bright AGNs (see \citealt{h2008} for a review). At first, the big blue bump disappears in LLAGNs \citep{s2005, h2009}. Secondly, \citet{x2011} and \citet{s2011} found that the optical-to-X-ray spectral index $\alpha_{\rm ox}$ (flux $F_\nu\propto \nu^{-\alpha_{\rm ox}}$, or equivalently, $\alpha_{\rm ox} = \log\left[L_\nu {\rm(2500\ \AA)}/L_\nu {\rm(2\ keV)}\right]/\log\left[\nu{\rm(2500\ \AA)}/\nu{\rm(2\ keV)}\right] = 0.384 \log\left[L_\nu {\rm(2500\ \AA)}/L_\nu {\rm(2\ keV)}\right]$.) correlates negatively with the bolometric Eddington ratio $\lambda$ in LLAGNs, but positively in bright AGNs. Thirdly, \citet{g2009} reported an anti-correlation between the hard 2-10 keV X-ray photon index $\Gamma$ and the Eddington ratio $\lambda$ in LLAGNs, in contrast with the positive correlation found in bright AGNs (see also \citealt{yx2015}). Fourthly, opposite correlations between the radio loudness and the luminosity ratio are also observed in LLAGNs and bright AGNs \citep{g2006}. Compared with black hole (BH) X-ray binaries (BHBs), all the above observations hint (see e.g., \citealt{g2006, h2008, s2011, yx2015}) that LLAGNs and bright AGNs are likely analogy to BHBs in their hard and soft states (see \citealt{b2010} for definitions of states in BHBs), respectively (but also see \citealt{d2014}).

From theoretical point of view, the accretion mode is different between LLAGNs (and BHBs in hard state) and bright AGNs (and BHBs in soft state). The BHBs in their soft state can be well described by an optically thick, geometrically thin accretion disc (\citealt{s1973}; the so-called Shakura-Sunyaev disc, SSD hereafter.), which emits thermal radiation. This thermal emission will appear in X-rays for BHBs in soft state. The SSD around supermassive BHs will be much cooler than that around stellar BHs (the case of BHBs), and they will emit in optical-to-ultraviolet (UV) bands. This will naturally explain the big blue bump commonly observed in bright AGNs. Additional observations also require a weak hot corona component, which contributes to the X-ray emissions. The BHBs in their hard state and the LLAGNs, on the other hand, will be better described by the truncated accretion--jet model (\citealt{e1997, l1996, y2003, w2007}. For reviews see \citealt{d2007, h2008, y2014}). In this model, the cold SSD is truncated at certain radius, within which it is replaced by a hot accretion flow. Besides, there is a relativistic jet near the BH. Generally the hot accretion flow is an update version of the conventional advection-dominated accretion flow (ADAF; \citealt{n1994, n1995, b1999, y2012, y2015, b2013, b2016, s2015}), and it is usually optically thin but geometrically thick (for review see \citealt{y2014}).

Given the above simple theoretical interpretation, further investigations are required, in order to understand observations in detail. For example, apart from the global trends observed, large dispersion is also witnessed in both the $\Gamma$--$\lambda$ correlation (e.g., \citealt{g2009,yx2015}) and the $\alpha_{\rm ox}$--$\lambda$ correlation (e.g., \citealt{x2011, s2011}). Obviously these large scatter indicates that, besides the accretion rate $\dot{M}$ which controls the luminosities in different wavebands, additional parameters/factors awaiting to be identified should also play certain roles. One possible factor is the difference in the viscosity parameter $\alpha$ of the hot accretion flow among different systems. Under a hot accretion flow model, systems with larger $\alpha$ can transport angular momentum more easily. Consequently, for a given accretion rate, these high-$\alpha$ systems will have higher infalling velocity and lower gas density, which results in lower broadband emission with larger $\alpha_{\rm ox}$ (\citealt{m1997,e1997}, and Section 4 for more discussions). Interestingly, based on local shearing-box magnetohydrodynamic (MHD) simulations of hot accretion flows, the ``effective'' viscosity parameter indeed can vary significantly, for a moderate change in the net magnetic flux and/or magnetic field configuration carried by the accreted material (e.g., \citealt{h1995, b2008, bs2013, s2016}). Besides, the viscosity parameter of its hot accretion component in AGN NGC 7213 can be well-constrained based on observations of a hybrid radio/X-ray correlation together with a V-shaped X-ray index/luminosity correlation \citep{x2016}, i.e. $\alpha\approx 0.01$, a value considerably smaller compared to the typical value of hot accretion flows ($\alpha\sim 0.1-0.3$, see e.g., \citealt{y2003, w2007, y2011}).

With the motivation to reveal the possible impact of viscosity parameter $\alpha$ (or more aggressively, the strength and/or configuration of magnetic field) of hot accretion flows, we in this work only consider quantities that relate to relative luminosities (i.e. Eddington ratios) instead of absolute ones, hoping to (partially) eliminate the impact of black hole mass. In practice, we focus on the optical-to-X-ray spectral index $\alpha_{\rm ox}$ and the radio-loudness $R_{\rm UV}$ in LLAGNs. Here parameter radio-loudness $R_{\rm UV}$ is defined as the ratio between the radio luminosity (spectral, at 5\ GHz, $L_{\rm R}=L_\nu {\rm (5\ GHz)}$) and UV luminosity (spectral, at 2500\ \AA, $L_{\rm UV}=L_\nu {\rm (2500\ \AA)}$), i.e. $R_{\rm UV}=L_{\rm R}/L_{\rm UV}$. The sample compilation is shown in Section 2. Since we target at hot accretion (e.g., ADAF) dominated systems, we thus limit ourselves to systems whose Eddington ratio $\lambda\la 10^{-3}$, an order of magnitude lower than the typical critical luminosity ($\lambda_{\rm crit}\sim 10^{-2}$) between bright AGNs and LLAGNs. We further exclude from our sample several faint sources whose X-rays may be of jet origin instead of hot accretion flow origin. Our final LLAGN sample includes 32 sources, among which 14 are low luminosity (in the sense of Eddington ratio $\lambda$) Seyfert galaxies and 18 are low-ionization nuclear emission-line region (LINER) galaxies. The Eddington ratio is in the range $10^{-6}\la \lambda \la 10^{-3}$. The results are shown in Section 3, where we report a strong negative correlation between $R_{\rm UV}$ and $\alpha_{\rm ox}$ in LLAGNs, independent of the luminosities. Interestingly, such correlation seems to be invisible based on a small sample of bright AGNs (See Section 3.2 for details.). Additional correlations, e.g. $R_{\rm UV}$--$\lambda$ and $\alpha_{\rm ox}$--$\lambda$ relationships are also reported. We then in Section 4 provide a comprehensive interpretation on the results (both the correlation and the scatter), based on the truncated accretion--jet model. Finally, Section 5 is denoted to a brief summary.

\section{The LLAGN SAMPLE with Eddington ratio $10^{-6} \la \lambda \la 10^{-3}$}\label{sample}

\begin{table*}
\normalsize
\centering
\caption{The sample of LLAGNs with $10^{-6}\la \lambda \la 10^{-3}$.}
\begin{minipage}{\textwidth}
\begin{center}
\begin{tabular}{lcccccccc}
\hline
\hline
{Name} & {log$M_{\rm
BH}/M_{\odot}$} &  {$\log L_{\rm R}$} &
{$\log L_{\rm UV}$} &  {$\log\lx$} &
{$\log \lambda$} &  {$\log R_{\rm UV}$} &  {$\log R_{\rm X}$} &  {$\alpha_{\rm ox}$}
\\
{(1)} &  {(2)} &  {(3)} &  {(4)} &
 {(5)} &  {(6)} &  {(7)} &  {(8)}  &  {(9)}\\
\hline
\multicolumn{9}{c}{\it Seyfert galaxies}\\
\hline
NGC 2639$^{(a)}$ &  8.02  &  29.52  & $<$24.18  &  22.80 & -3.82  & $>$5.34  & 6.72& $<$0.53\\
NGC 4138$^{(a)}$ &  7.75  &  28.98  & $<$23.94  &  23.21 & -3.08  & $>$5.04  & 5.77& $<$0.28\\
NGC 4168$^{(a)}$ &  7.95  &  27.88  & $<$22.97  &  21.98 & -4.70  & $>$4.91  & 5.90& $<$0.38\\
NGC 4258$^{(a)}$ &  7.61  &  26.25  & $<$23.36  &  22.75 & -3.37  & $>$2.89  & 3.5 & $<$0.24\\
NGC 4565$^{(a)}$ &  7.70  &  27.22  &    24.79  &  21.47 & -4.89  &    2.43  & 5.75&  1.28\\
NGC 4579$^{(a)}$ &  7.78  &  28.50  &    25.55  &  23.10 & -3.37  &    2.95  & 5.40&    0.94\\
NGC 4639$^{(a)}$ &  6.85  &  27.65$^\ast$  &    24.19  &  22.26 & -3.25  &    3.46 &    5.39 & 0.74\\
NGC 2685$^{(a)}$ &  7.15  &  26.07  &    23.96  &  21.42 & -3.82  &    2.11  & 4.65&    0.97\\
NGC 3147$^{(a)}$ &  8.79  &  28.23  &    25.12  &  23.96 & -3.52  &    3.11  & 4.27&    0.45 \\
NGC 3486$^{(a)}$ &  6.14  &  28.13  &    23.06  &  20.53 & -3.89  &    5.07 & 7.60&    0.97 \\
NGC 4477$^{(a)}$ &  7.92  &  27.55$^\ast$  &    24.29  &  21.72 & -4.89  &    3.26  & 5.83&    0.99\\
NGC 4501$^{(a)}$ &  7.90  &  27.60  &    24.79  &  21.51 & -4.92  &    2.81 & 6.09&    1.26 \\
NGC 4698$^{(a)}$ &  7.84  &  27.09$^\ast$  &    23.74  &  21.27 & -5.30  &    3.35 & 5.82&    0.95 \\
NGC 4725$^{(a)}$ &  7.49  &  28.12$^\ast$  &    23.53  &  20.96 & -5.22  &    4.59 & 7.16&    0.99 \\
\hline
\multicolumn{9}{c}{\it LINERs}\\
\hline
NGC  266$^{(a)}$  &  7.90  &  28.22  &    25.50  &  22.76 & -3.64  &    2.72 & 5.46&    1.05 \\
NGC 0315$^{(a)}$  &  9.24  &  30.47  &    25.55  &  23.56 & -4.22  &    4.92 & 6.91&    0.77 \\
NGC 2681$^{(a)}$  &  7.20  &  27.48$^\ast$  &    24.10  &  20.95 & -4.88  &    3.88  & 6.53&    1.21\\
NGC 3718$^{(a)}$  &  7.97  &  27.34  &    23.83  &  21.97 & -4.53  &    3.51 & 5.37&    0.71 \\
NGC 4143$^{(a)}$  &  8.31  &  28.31  &    24.67  &  22.03 & -4.89  &    3.64 & 6.28&    1.02 \\
NGC 4278$^{(a)}$  &  9.20  &  28.54  &    24.72  &  21.94 & -5.86  &    3.82 & 6.60&    1.07 \\
NGC 4261$^{(a)}$  &  8.94  &  29.76  &    24.81  &  22.73 & -4.41  &    4.95 & 7.03&    0.80 \\
NGC 4457$^{(a)}$  &  7.00  &  27.71$^\ast$  &    25.00  &  20.99 & -4.63  &    2.71 & 6.72&    1.54 \\
NGC 4494$^{(a)}$  &  7.60  &  26.83$^\ast$  &    23.88  &  21.04 & -5.22  &    2.95 & 5.79&    1.09 \\
NGC 4548$^{(a)}$  &  7.51  &  26.92$^\ast$  &    23.43  &  21.79 & -4.34  &    3.49 & 5.13&    0.63 \\
NGC 4736$^{(a)}$  &  7.42  &  26.40  &    23.25  &  20.76 & -5.39  &    3.15 & 5.64&    0.96 \\
NGC 5746$^{(a)}$  &  7.49  &  29.02$^\ast$  &    23.55  &  21.88 & -4.04  &    5.47 & 7.14&    0.64 \\
NGC 6240$^{(a)}$  &  9.11  &  29.73  &    27.73  &  23.77 & -3.69  &    2.00 & 5.96&    1.52 \\
NGC  404$^{(b)}$  &  5.30  &$<$24.74 &    24.67  &  19.46 & -4.84  & $<$0.08 & $<$5.28&    2.00 \\
NGC 1052$^{(b)}$  &  8.10  &  29.98  &    25.16  &  22.86 & -4.12  &    4.81 & 7.12&    0.89 \\
NGC 3386$^{(b)}$  &  7.40  &$<$26.01 &    24.87  &  21.38 & -5.02  & $<$1.14 & $<$4.63&    1.34 \\
NGC 3642$^{(b)}$  &  7.10  &$<$26.65 &    25.70  &  22.61 & -3.49  & $<$0.96 & $<$4.04&    1.19 \\
NGC 4203$^{(b)}$  &  7.00  &  27.55  &    25.56  &  22.67 & -3.33  &    1.79 & 4.68&    1.11\\
\hline
\end{tabular}
\end{center}
\end{minipage}
Notes: Col.(1): Source name. Here the superscripts $^{(a)}$ and $^{(b)}$ represent sources that are taken from \citet{x2011} and \citet{m2007}, repectively. Col.(2): Black hole mass.  Col.(3): radio spectral luminosity at 5\ GHz, $L_{\rm R}$, in unit of ${\rm erg\ s^{-1}\ Hz^{-1}}$. Col.(4): UV spectral luminosity at 2500\ \AA, $L_{\rm UV}$, in unit of ${\rm erg\ s^{-1}\ Hz^{-1}}$. Col.(5): X-ray spectral luminosity at 2\ keV, $\lx$, in unit of ${\rm erg\ s^{-1}\ Hz^{-1}}$. Col.(6): Eddington ratio, $\lambda=\lbol/\ledd$. Col.(7): Radio loudness $R_{\rm UV}\equiv L_{\rm R}/L_{\rm UV}$. Col.(8): Radio loudness defined as, $R_{\rm X}=L_{\rm R}/L_{\rm X}$. Col.(9): optical-to-X-ray spectral index $\alpha_{\rm ox}\equiv 0.384 \log\left[L_{\rm UV}/\lx\right]$. \\
$^\ast$: For sources labelled with $^\ast$, their radio flux at 5 GHz are derived based on observations at neighbouring frequencies.
\vspace{0.1cm}
\end{table*}

\begin{table*}
\normalsize
\centering
\caption{The sample of bright AGNs with $\lambda > 10^{-2}$.}
\begin{minipage}{\textwidth}
\begin{center}
\begin{tabular}{lcccccccc}
\hline
\hline
{Name} & {log$M_{\rm
BH}/M_{\odot}$} &  {log$L_{\rm R}$} &
{log$L_{\rm UV}$} &  {log$L_{\rm X}$} &
{log$\lambda$} &  {log$R_{\rm UV}$} &  {$\log R_{\rm X}$}  &  {$\alpha_{\rm ox}$} \\
{(1)} &  {(2)} &  {(3)} &  {(4)} &
 {(5)} &  {(6)} &  {(7)} &  {(8)}  &  {(9)} \\
\hline
\multicolumn{9}{c}{\it Seyfert galaxies}\\
\hline
NGC 3516$^{(a)}$ &  7.36  &  27.55  &    27.14  &  24.14 & -1.70  &    0.41 &   3.41&    1.15 \\
NGC 4051$^{(a)}$ &  6.11  &  27.26  &    25.72  &  23.16 & -1.42  &    1.54 &   4.10&    0.98 \\
NGC 4151$^{(a)}$ &  7.18  &  28.64  &    27.56  &  24.31 & -1.33  &    1.08 &   4.33&    1.25 \\
NGC 4388$^{(a)}$ &  6.80  &  28.81  &    26.14  &  23.72 & -1.70  &    2.67 &   5.09&    0.93 \\
NGC 4395$^{(a)}$ &  5.04  &  25.59  &    24.16  &  21.65 & -1.85  &    1.43 &   3.94&    0.96 \\
NGC 5273$^{(a)}$ &  6.51  &  26.86  &    25.53  &  23.27 & -1.77  &    1.33 &   3.59&    0.87 \\
NGC 5548$^{(a)}$ &  8.03  &  28.87  &    27.16  &  25.25 & -1.40  &    1.71 &   3.62&    0.74 \\
\hline
\end{tabular}
\end{center}
\end{minipage}
Notes: Col.(1)-(9) are the same as those in Table 1.
\end{table*}

We gather from literature a sample of LLAGNs, which consist of both LINERs and low-$\lambda$ Seyfert galaxies. Firstly, we collect the nuclear emission of 16 low-luminosity Seyfert galaxies from \citet{x2011}. All the objects in \citet{x2011} were gathered by \citet{g2009} from the Palomar sample \citep{h1997a,h2008} and the multi-wavelength catalog of LINERs \citep{c1999}, where all the sources have \textit{Chandra} and/or \textit{XMM-Newton} observations. \citet{x2011} further excluded six LINERs who lack estimation in UV luminoisties. The optical luminosity at 2500\ \AA\ is derived from the absolute B magnitude mainly observed by high-spatial-resolution {\it Hubble Space Telescope} (HST, \citealt{h1997a,h1997b}), with the assumption that the optical-UV spectral index is $\alpha_{o} = -1$, a typical value for the optical featureless continuum of Seyfert 1 nuclei \citep{w1987}, cf. \citet{x2011} for details. For the core emission of LINERs, we gather from \citet{m2007} and \citet{x2011}, which includes respectively, 13 and 20 sources. The data compilation of LINERs is the same to that of Seyferts in \citet{x2011}. The sample of \citet{m2007} are from \citet{m2005}, where four objects are excluded for different reasons (see \citealt{m2007} for details). All the objects in \citet{m2005} are optically classified as LINERs by \citet{h1997a}, and they are selected provided their UV is observed by high-resolution HST and their nuclei X-ray emission is from {\it Chandra} (or {\it XMM-Newton}). Considering the duplicates in the two works, there are 27 LINERs left in total. We note that there are other LINER samples in literature, e.g., \citet{yo2011}, \citet{h2013}, and \citet{c2016}, and all these samples are also selected from \citet{h1997a,h1997b} and/or \citet{c1999}. Consequently, except for several sources that lack constraints in X-ray photon index $\Gamma$, most of them overlap with the samples of \citet{m2007} and \citet{x2011}.

Additionally, there are faint LLAGNs whose X-rays may be dominated by emission from relativistic jets rather than hot accretion flows (e.g., \citealt{y2009, n2010, n2014, x2017}), and these sources are not justified for our scientific motivation (to investigate the properties of a hot accretion flow itself), thus should be excluded from our sample. From theoretical point of view, it is because, as the accretion rate reduces, the relative importance of jet emission increases \citep{y2005}. Below a certain critical luminosity (statistically the critical luminosity is $L_{\rm X}{\rm(2-10\ keV)} \sim 10^{-6}\ \ledd$; \citealt{y2005, y2009, x2017}), the X-ray emission from jets will exceed that of hot accretion flows. Six sources (NGC3031, NGC3998, NGC4374, NGC4486, NCG4552, NGC4594) are excluded from our sample, as they have confirmation of the jet origin in X-rays from multi-waveband spectral modelling \citep{w2007,y2009,y2011,n2014}. Besides, The other five sources (NGC3941, NGC3226, NGC3169, NGC6500, and NGC7130) are also excluded from our sample, for the reason that their X-ray spectra are relatively soft, with photon index $\Gamma > 2$ \citep{m2007, g2009}, which is suggestive of a jet origin in X-rays considering their faintness in X-rays \citep{c2014, n2014, yx2015}.

As listed in Table 1, our final LLAGN sample includes 14 low-$\lambda$ Seyfert galaxies and 18 LINERs. Sources with superscript $^{(a)}$ in Table 1 are gathered from \citet{x2011}. For these sources, Cols.(1), (2), (4), (5), (6) and (9) are directly taken from \citet{x2011}. We additionally gather its nuclear radio luminosity at 5 GHz from NED,\footnote{website: http://ned.ipac.caltech.edu/} as shown in Col. (3). For sources without radio observations at 5 GHz (labelled with superscript $^\ast$), we derive its spectral radio luminosity at 5\ GHz based on observations at neighbouring frequencies, with the spectral index in radio wavebands assumed to be $\alpha_{\rm r}\approx0$. This choice of $\alpha_{\rm r}$ comes from the fact that the radio spectra of LLAGNs tend to be flat or even slightly reverted (see, e.g., \citealt{u2001,h2008}), a signature of self-absorbed synchrotron emission from conical relativistic ``continuous'' jet \citep{b1979}. Such derivation of $L_{\rm R}$ should not introduce large scatter to the correlations reported later in this work, since we usually have $-0.3<\alpha_{\rm r} <0.3$ for LLAGNs \citep{u2001, h2008}, and the adopted frequency is close to 5\ GHz. With $L_{\rm R}$, we subsequently can calculate the radio loudnesses $R_{\rm UV}$ and $R_{\rm X}$ (here we define $R_{\rm X}=L_{\rm R}/L_{\rm X}$, following \citealt{h2008}.), respectively, which are shown in Cols.(7) \& (8).

\begin{figure}
\centering
\includegraphics[width=0.5\textwidth]{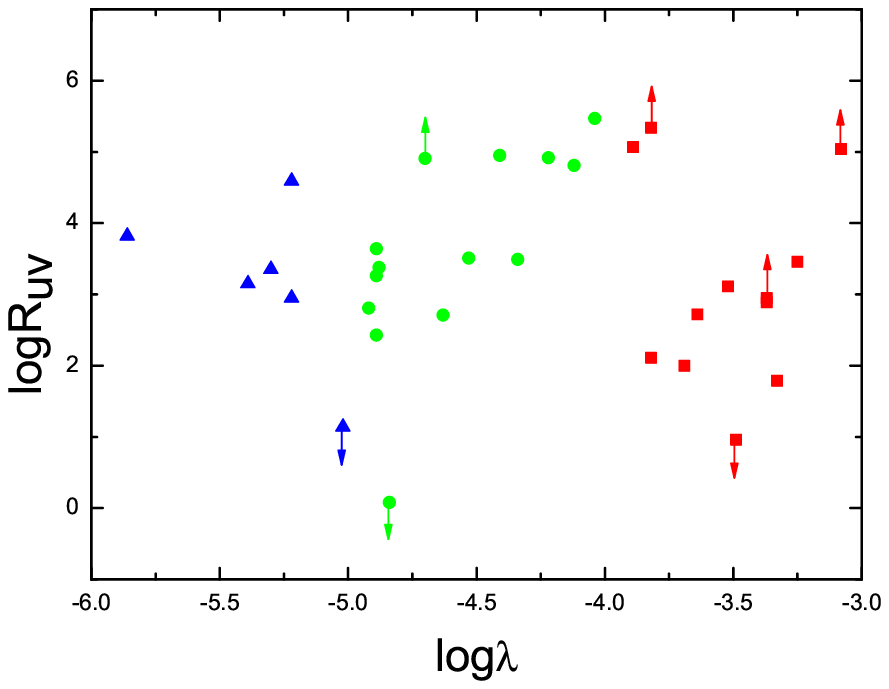}
\vspace{-.5cm}
\includegraphics[width=0.5\textwidth]{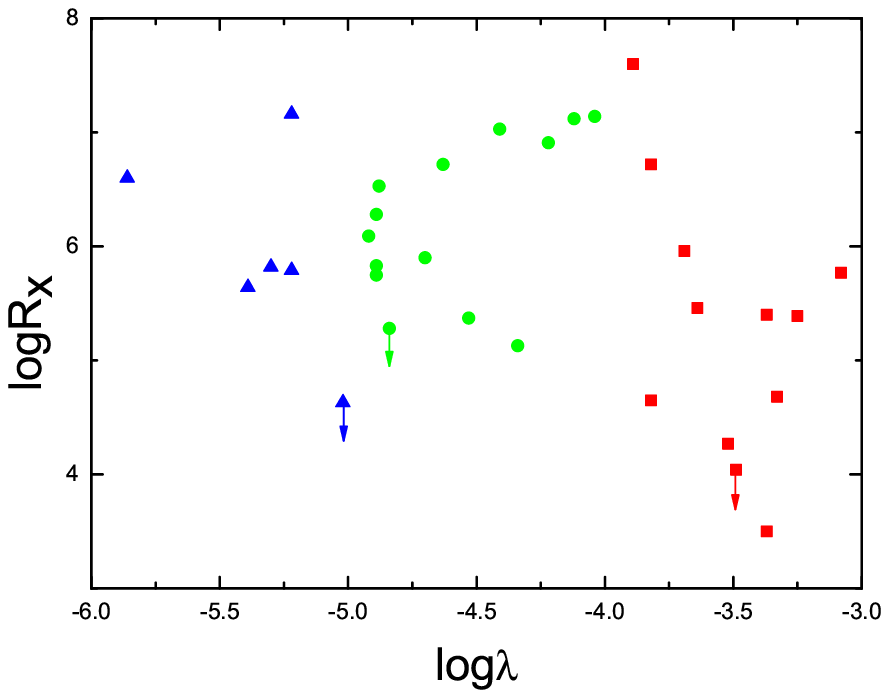}
\vspace{-.5cm}
\includegraphics[width=0.5\textwidth]{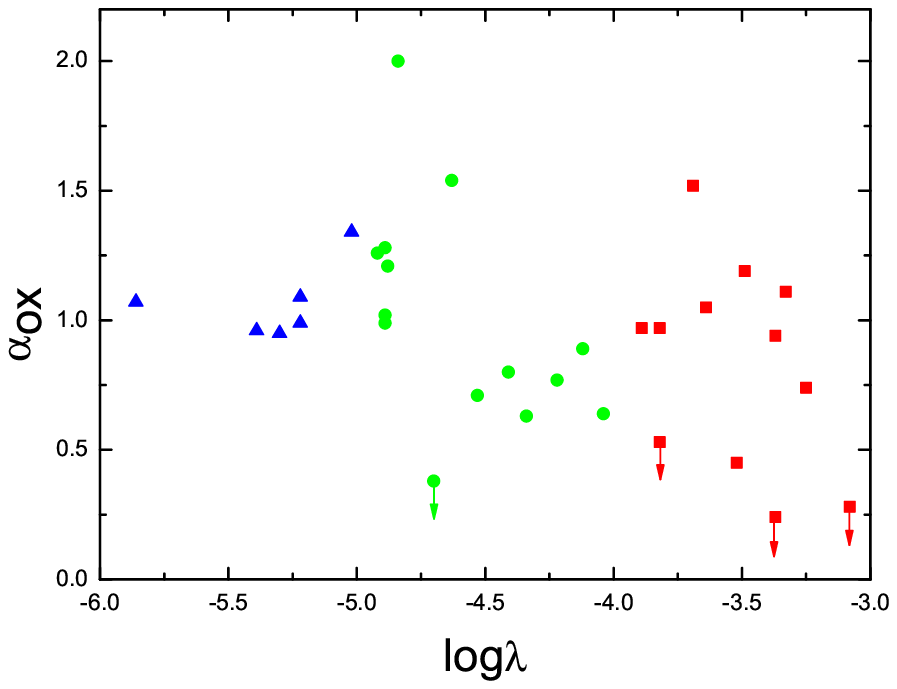}
\vspace{-0.5cm}\caption{Radio loudnesses ($R_{\rm UV}$:{\it Top panel}; $R_{\rm X}$: {\it Middle panel}) and optical/UV-to-X-ray spectral index $\alpha_{\rm ox}$ ({\it bottom panel}), respectively, as a function of the Eddington ratio $\lambda$ for LLAGNs. The blue triangles, green circles and red squares are, respectively, for sources whose Eddington ratio is in the range of $-6 < \log\lambda < -5$, $-5 < \log\lambda < -4$, and $-4 < \log\lambda < -3$. }\label{llagnsample}
\end{figure}

\begin{figure*}
\vspace{-1cm}
\includegraphics[width=0.7\textwidth]{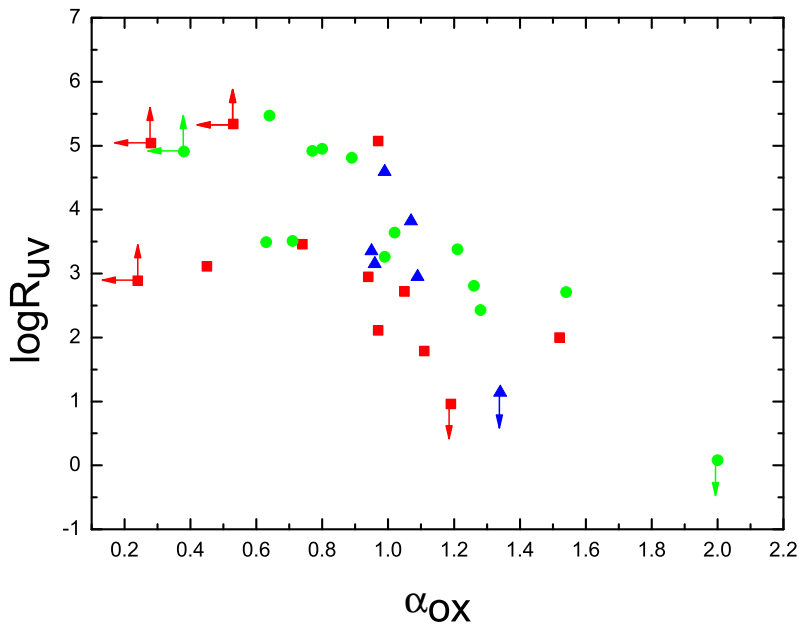}
\vspace{-0.8cm}\caption{Radio loudnesses $R_{\rm UV}$ as a function of the optical/UV to X-ray spectral index $\alpha_{\rm ox}$ for LLAGNs. The symbols are the same to those in Fig.\ \ref{llagnsample}.}\label{r-ox}
\end{figure*}

For sources with superscript $^{(b)}$ in Table 1, all data except Col.(6) and Col.(8) are adopted from \citet{m2007} directly. The Eddington ratio in Col. (6) is estimated based on its 2-10\ keV X-ray luminosity $L_{\rm X} {\rm (2-10\ keV)}$, i.e. we assume the bolometric luminosity $\lbol\approx30\ L_{\rm X} {\rm (2-10\ keV)}$, the same as that in \citet{x2011}. Note that another correction $\lbol\approx 16\ L_{\rm X} {\rm (2-10\ keV)}$, which is approximately a factor of $\sim$2 lower than our choice, is also widely adopted in literature (see e.g., \citealt{h2008}). Since our results are insensitive to the exact bolometric luminosity of each source, we omit further discussions on the two corrections below.

\section{RESULTS AND DISCUSSIONS}\label{RESULTS}

\subsection{basic properties of the LLAGN sample}

Before we investigate the relationship between $R_{\rm UV}$ and $\alpha_{\rm ox}$, we first show in Fig.\ \ref{llagnsample} the general properties of our sample. We divide our sample into three luminosity regimes, i.e. $-6 < \log\lambda < -5$ (blue triangles in all the plots), $-5 < \log\lambda < -4$ (green circles) and $-4 < \log\lambda < -3$ (red squares).

From the top panel of Fig.\ \ref{llagnsample}, a liner fit (in logarithmic space) between $R_{\rm UV}$ and $\lambda$ reads,
\begin{equation}
\log R_{\rm UV}=(0.04\pm0.33)\log\lambda+3.51\pm1.44,\label{eq:ruvl}
\end{equation}
where the confidence level of rejecting a null hypothesis is $9.4\ \%$, based on Pearson test. We note that based on larger samples that have sufficiently large dynamical range in $\lambda$ (especially the bright end), we will normally observe a negative $R_{\rm UV}$--$\lambda$ correlation \citep{m2007, s2007}. Since the null (or weak positive) $R_{\rm UV}$--$\lambda$ correlation observed in our work suffers large uncertainties in the correlation slope, we will rely on those previous works later for the theoretical interpretation.

The relationship between $R_{\rm X}$ and $\lambda$, on the other hand, is much stronger (cf. the middle panel of Fig.\ \ref{llagnsample}), i.e. a liner fit in logarithmic space gives,
\begin{equation}
\log R_{\rm X}=(-0.46\pm0.23)\log\lambda+3.83\pm1.00, \label{eq:rlambda}
\end{equation}
whose confidence level is $94.7\ \%$. This negative $R_{\rm X}$--$\lambda$ correlation had been reported in numerous works (cf., \citealt{h2008} and references therein)\footnote{Note that a rough estimation of the LLAGN sample of \citet{h2008}, which has a larger dynamical range in $\lambda$ ($\sim$ 8 orders of magnitude), indicates $\log R_{\rm X}\sim -0.8\log\lambda+const.$. But this steep correlation may be contaminated by AGNs at bright end. When limited to sources with $10^{-6}< \lambda < 10^{-3}$, the slope becomes shallower (see their Fig. 10, and also Fig. 3 in \citealt{s2007}).}, as LLAGNs are systematically louder in radio compared with bright AGNs.

The bottom panel of Fig. \ref{llagnsample} shows the relationship between optical/UV-to-X-ray spectral index $\alpha_{\rm ox}$ and Eddington ratio $\lambda$. A negative correlation can be derived, i.e. from linear fitting it reads,
\begin{equation}
\alpha_{\rm ox}=(-0.19\pm0.09)\log\lambda+0.12\pm0.37.
\end{equation}
The confidence level of this result is high, $96.8\ \%$. This result is consistent with that given in \citet{x2011}, where the correlation slope is reported to be $-0.16$.


\subsection{the $R_{\rm UV}$--$\alpha_{\rm ox}$  relationship in LLAGNs}

We now investigate the relationship between radio-loudness and optical/UV-to-X-ray spectral property. As shown in Fig.\ \ref{r-ox}, it is evident that there is a negative correlation between $R_{\rm UV}$ and $\alpha_{\rm ox}$ for all the three luminosity regimes. The confidence level of correlation analysis for the three regimes are, respectively, $95.4\ \%$ (the $-4<\log\lambda<-3$ regime), $99.9\ \%$ (the $-5<\log\lambda<-4$ regime) and $95.3\ \%$ (the $-6<\log\lambda<-5$ regime). For the whole sample, the confidence level of a negative correlation is larger than $99.9\ \%$, remarkably strong. A linear fit ($R_{\rm UV}$ in logarithmic space) to the data shows that,
\begin{equation}
\log R_{\rm UV} = (-2.39\pm 0.48)\ \alpha_{\rm ox} + 5.61\pm 0.49 \label{eq:ruvobs}
\end{equation}

\begin{figure}
\centering
\vspace{-0.5cm}\includegraphics[width=8.5cm]{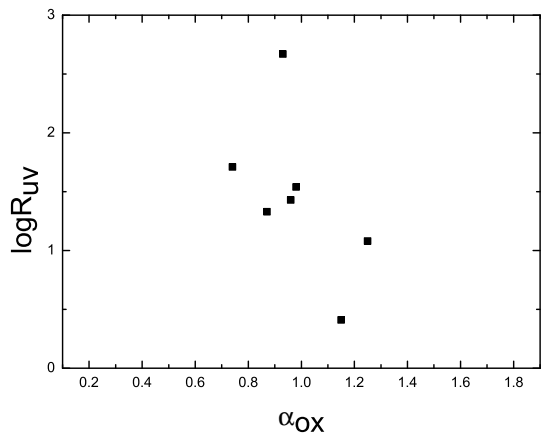}
\vspace{-1cm}\caption{The relationship between radio loudness $R_{\rm UV}$ and the optical/UV to X-ray spectral index $\alpha_{\rm ox}$ for bright AGNs with $\lambda>10^{-2}$. }\label{r-ox-agn}
\end{figure}

For comparison, we show in Fig.\ \ref{r-ox-agn} the relationship between $R_{\rm UV}$ and $\alpha_{\rm ox}$ for a limited sample of bright AGNs, i.e. Seyfert galaxies with $\lambda > 10^{-2}$. These sources, mainly gathered from \citet{x2011}, are listed in Table 2. From this figure we find that the correlation between $R_{\rm UV}$ and $\alpha_{\rm ox}$ in bright AGNs (or at least bright Seyferts) is absent/weak, at the confidence level $80.3\ \%$. We note that this bright AGN sample is obviously incomplete; there are numerous bright AGNs exist in literature. However, since in this work we intend to focus on LLAGNs only, the expansion of bright AGNs will be devoted to a separate work.

\section{theoretical interpretation based on truncated accretion--jet model}

It has been known for years that, in LLAGNs systems, there exist negative correlations between radio-loudness $R_{\rm UV}$ and Eddington ratio $\lambda$ (or UV/X-ray Eddington ratios. See e.g., \citealt{m2007, h2008}. Admittedly the LLAGN sample in this work provides a null correlation but with large uncertainties, cf. Eq.\ \ref{eq:ruvl}.), and between optical-to-X-ray spectral index $\alpha_{\rm ox}$ and Eddington ratio $\lambda$ \citep{x2011}. We in this work, report a new negative correlation, i.e. the $R_{\rm UV}$--$\alpha_{\rm ox}$ correlation, which is likely insensitive to the Eddington ratio. We emphasis that all the three correlations witness large scatter, i.e., for a given $\lambda$, both $R_{\rm UV}$ and $\alpha_{\rm ox}$ vary significantly from one source to the other.

We in this Section will try to understand the three correlations as well as their scatter under the truncated accretion--jet model, a model that has been successfully applied to LLAGNs (for review see \citealt{y2014}), e.g., to understand their broadband spectra and also the ``fundamental plane'' of BH activity. The accretion--jet model utilizes three components (for details, cf. \citealt{e1997, y2014}), i.e. an outer cold SSD, which is truncated at certain radius, an inner hot accretion flow within that radius, and a conical jet near the central black hole. In this work, we argue that the scatters themselves in the three correlations provide additional information on the fundamental properties of accretion flows. Before detailed theoretical analysis, we first speculate that the relativistic beaming effect is not the dominated reason for the scatter in radio-loudness. This is mainly because we exclude in our sample sources with small-viewing-angles, such as Blazars. If the jet is viewed at moderately large (and similar in value) viewing angles, then the difference in the beaming enhancement should be small among different sources \citep{c2017}. Consequently, the scatter in radio emission (or $R_{\rm UV}$) is mainly intrinsic.

We emphasis that, if $R_{\rm UV}$ and $\alpha_{\rm OX}$ depend solely on the accretion rate (or equivalently the luminosity $\lambda$), i.e. $R_{\rm UV}\propto \lambda^{-\xi_1}$ and $\alpha_{\rm OX}\propto \lambda^{-\xi_2}$ (here $\xi_1>0, \xi_2>0$), then their negative correlations will result in a positive $R_{\rm UV}$--$\alpha_{\rm OX}$ relationship, i.e. $R_{\rm UV}\propto\alpha_{\rm OX}^{\xi_1/\xi_2}$, contradict to what we observed. Even for the weak positive $R_{\rm UV}$--$\lambda$ correlation reported from our sample, the correlation coefficients $\xi_1 = -0.04$ and $\xi_2 = 0.46$ will lead to a weak relationship, i.e., $R_{\rm UV}\propto\alpha_{\rm OX}^{-0.09}$, inconsistent with the steep one shown in Eq.\ \ref{eq:ruvobs} (see also Fig.\ \ref{r-ox}). It is thus clear that the observed strong negative $R_{\rm UV}$--$\alpha_{\rm OX}$ correlation should be caused by other factors. Our key idea to this problem, as well as the scatter observed, is to consider the effect of viscosity parameter $\alpha$.

\subsection{basic properties of truncated accretion--jet model}

According to hot accretion flow theory \citep{n1994,y2014}, larger $\alpha$ will lead to higher radial infalling velocity $V_R$, as $V_R \sim -\alpha C_s^2/\Omega_K$, where $C_s$ is the sound speed (insensitive to $\dot{m}$ when radiative cooling is dynamically unimportant, the case explored in this work.) and $\Omega_K$ is the Keplerian rotational velocity \citep{y2014}. Consequently the density $n$ and the surface density $\Sigma$ of the accretion flow will be lower, i.e. $n\propto \dot{m}/\alpha$ and $\Sigma\propto \dot{m}/\alpha$. Hot accretion flow is optically thin, with synchrotron, bremsstrahlung, and the Compton scattering as its major radiation mechanisms \citep{n1995}. The bolometric luminosity can then be given as,
\begin{equation}
\lbol\sim (\frac{\dot{m}}{\alpha})^a, \label{eq:lbol}
\end{equation}
where the index $a$ is estimated to be, $a\approx 2$, based on detailed numerical calculations of hot accretion flows \citep{e1997,m2003,x2012}.

For LLAGNs, there are two competing components for the emission at optical/UV wavebands. One component is the thermal emission from the outer truncated SSD, which peaks in optical or UV bands, depending on the location of truncated radius. The other component is the Compton scattering of synchrotron emission from the inner hot accretion flow \citep{m1997,y2014}. Since our sample is limited to sources whose $\lambda<10^{-3}$, we argue that the emission at 2500\ \AA\ mainly comes from the inner hot accretion flow\footnote{Obviously, for brighter systems, i.e. those with $10^{-3}\la\lambda\la10^{-2}$, the truncated radius is small enough that the emission from the outer SSD should be taken into account.}. Interestingly, for hot accretion flows around supermassive BHs (the LLAGN case), the optical/UV emission usually is the first Compton up-scattered bump \citep{m1997,y2014}. The emission from such process has a positive correlation with $\dot{m}$, as $\dot{m}$ will determine mainly the optical depth (surface density), a quantity that controls the probability of Compton scattering \citep{s1980, d1991}. On the other hand, the first Compton bump is even more sensitive to the energy of electrons (a.k.a. the electron temperature), as it determines the location (in wavebands) of the peak. For a given $\dot{m}$, larger $\alpha$ means lower density, or equivalently lower radiative cooling to the electrons. Consequently, the electrons will be more energetic with higher temperatures \citep{x2012}. The dependence of emission at optical/UV band on parameter $\alpha$ relies on detailed numerical calculations, where \citet{m1997} reported that the emission at 2500\ {\AA} has a positive correlation with $\alpha$. With the above reasons, we simply write $L_{\rm UV}$ as,
\begin{equation}
L_{\rm UV}\sim (\dot{m} \alpha)^b,
\end{equation}
where $b\sim2$. The estimation of $b$ comes from the fact that, the first-order Compton scattering (responsible for the optical/UV emission) depends on the product of the seed photon flux (synchrotron emission, $\propto\dot{m}^{0.5-1}$, cf. \citealt{mah1997,y2015}, note that electron temperature also depends on $\dot{m}$ for the expression in \citealt{mah1997}.) and optical depth in vertical direction ($\propto\dot{m}^{1-2}$).

The X-ray emission mainly comes from the high-order (second or even higher) Compton scattering processes in hot accretion flow, which is obviously more sensitive to the optical depth \citep{s1980, d1991}, compared to the first-order Compton bump (e.g., $L_{\rm UV}$). With the assumption that the optical depth is relatively small in hot accretion flows, we may write it as,
\begin{eqnarray}
\lx & \propto & L_{\rm UV}\, \alpha^{-2}\, \Sigma^c \nonumber\\
 & \propto & \dot{m}^{b+1} \alpha^{b-3},
\end{eqnarray}
where the index $c\approx 1$--2. Because we exclude moderately bright LLAGNs (whose $10^{-3} < \lambda < 10^{-2}$) from our sample, we may safely assume $c\approx 1$ in the second expression. In the above expression, the factor $\alpha^{-2}$ is introduced to take into account the shift of photon energy (in $h\nu$) of both synchrotron emission and consequently the first Compton bump, as viscosity parameter $\alpha$ changes (cf. \citealt{m1997} for numerical calculations). We also note that, from statistical point of view, the integrated (in 2-10\ keV energy band) X-ray luminosity of LLAGNs can roughly be expressed as $\lx{\rm (2-10\ keV)} \approx f_{\rm X}\ \lbol$, where observationally the correction factor $f_{\rm X}$ is estimated to be $f_{\rm X}\approx 1/30$ \citep{g2009, x2011} or $f_{\rm X}\approx 1/16$ \citep{h2008}.

Now we come to estimate the self-absorbed synchrotron radio emission, whose $\alpha_{\rm r}\approx 0$, from the jet. Based on conical ``disc--jet'' model, the jet is formed by extracting the rotation energy of the underlying disc. Moreover, the composition of the disc--jet is assumed to be dominated by normal plasma, i.e. electrons and ions, coming from the underlying hot accretion flow. The flat-spectrum radio emission (at e.g., 5 GHz) has a power-law dependence on the mass loss rate into the jet $\dot{m}_{\rm jet}$ \citep{hs2003, g2014}, i.e., we have,
\begin{eqnarray}
L_{\rm R}& \propto & \dot{m}_{\rm jet}^{ 1.4}\nonumber\\
& \propto & \dot{m}^{1.4},
\end{eqnarray}
where in the latter expression we assume $\dot{m}_{\rm jet}\propto \dot{m}$. Note that detailed modelling on observations implies a slightly weaker dependence on $\dot{m}$ (e.g., \citealt{y2005,xy2016}). When this correction is taken into account, we would observe slightly steeper correlations between $R_{\rm UV}$  and $\lambda$, and $R_{\rm X}$ and $\lambda$, while there is no impact to $\alpha_{\rm ox}$, see below.

Therefore, the radio loudnesses $R_{\rm UV}$  and $R_{\rm X}$ can be expressed as,
\begin{eqnarray}
R_{\rm UV} & = &\frac{L_{\rm R}}{L_{\rm UV}}\propto \dot{m}^{1.4-b}\alpha^{-b}\propto (\dot{m}/\alpha)^{1.4-b} \alpha^{1.4-2b}, \label{eq:ruv}\\
R_{\rm X} & = & \frac{L_{\rm R}}{L_{\rm X}}\propto \dot{m}^{0.4-b}\alpha^{3-b}\propto(\dot{m}/\alpha)^{0.4-b} \alpha^{3.4-2b}.
\label{eq:rx}
\end{eqnarray}
The optical-to-X-ray spectral index $\alpha_{\rm ox}$ can be written as,
\begin{equation}
\alpha_{\rm ox} = 0.384\log (L_{\rm UV}/\lx) \approx -0.4\log(\dot{m}/\alpha) + 0.8\log\alpha  +const.\label{eq:alphaox}
\end{equation}

\subsection{theoretical interpretation on three correlations and their scatters}
We now apply the above results to understand the three correlations observed. If we assume that LLAGNs on average have a mean value of $\alpha$, $\bar{\alpha}$. Then substitute Eq.\ \ref{eq:lbol} to Eqs.\ \ref{eq:ruv}--\ref{eq:alphaox}, we have,
\begin{eqnarray}
\log R_{\rm UV} & \approx & -{(b-1.4)\over a}\log\lbol - (2b-1.4)\log\bar{\alpha} + const.\nonumber\\
& \approx & -0.3\log\lambda -2.6\log\bar{\alpha} + const.,\label{eq:ruv1}\\
\log R_{\rm X} & \approx & -{(b-0.4)\over a}\log\lbol - (2b-3.4)\log\bar{\alpha} + const.\nonumber\\
& \approx & -0.8\log\lambda -0.6\log\bar{\alpha} + const.,\label{eq:rx1}\\
\alpha_{\rm ox} & \approx & -(0.4/a)\log\lbol+ 0.8\log\bar{\alpha}+ const.\nonumber\\
& \approx & -0.2\log\lambda + 0.8\log\bar{\alpha}+const.\label{eq:alphaox1}
\end{eqnarray}
For all the second expressions, we take $a\approx2$ and $b\approx 2$.

Obviously, based on the truncated accretion--jet scenario, we have naturally reproduced negative correlations between $R_{\rm UV}$ and $\lambda$ (Eq.\ \ref{eq:ruv1}, for observational results see Sec. 3 and \citealt{m2007}, \citealt{s2007}), $R_{\rm X}$ and $\lambda$ (Eq.\ \ref{eq:rx1}, for observational results see Sec. 3 and \citealt{h2008}), and $\alpha_{\rm ox}$ and $\lambda$ (Eq.\ \ref{eq:alphaox1}, for observational results see Sec. 3 and \citealt{x2011}). How to understand the null relationship between $R_{\rm UV}$ and $\lambda$ reported in this work (but inconsistent with earlier works)? From Eq.\ \ref{eq:ruv1}, we argue it is likely because that $R_{\rm UV}$--$\lambda$ correlation is relatively weak (with a slope of -0.3) compared to the large scatter introduced by $\alpha$ (with a slope of -2.6). Interestingly, as noted before in footnote 2, the $R_{\rm UV}$/$R_{\rm X}$--$\lambda$ correlations will be shallower when we limit to sources with $10^{-6} \leq \lambda \leq 10^{-3}$ (Fig. 3 in \citealt{s2007} and Fig. 10 in \citealt{h2008}). The correlation slope derived from the whole AGN sample may be contaminated by those brighter end AGNs. The scatter observed in the three correlations, on the other hand, could be considered as the deviations of $\alpha$ of individual sources to the mean value $\bar{\alpha}$. We note that \citet{x2011} had already reported a negative $\alpha_{\rm ox}$--$\lambda$ correlation together with detailed numerical calculations of hot accretion flow.

There are two points in our theoretical interpretation that worth further emphasis. The first is that, the $R_{\rm X}$--$\lambda$ correlation is steeper than the $R_{\rm UV}$--$\lambda$ correlation. The second is that, $R_{\rm X}$--$\lambda$ correlation has a much weaker dependence on $\alpha$, compared to the $R_{\rm UV}$--$\lambda$ correlation. Combining these two reasons, we expect that, if $\alpha$ varies from source to source, then the $R_{\rm UV}$--$\lambda$ relationship will have larger scatter.

Then, what determines the value of $\alpha$ in accretion flows? According to recent advanced MHD simulations \citep{bs2013,s2016}, the value of ``effective'' $\alpha$ varies from 0.01 to 1, depending on the net vertical magnetic flux and/or magnetic field configuration carried by the accreted material. Interestingly, such large difference in $\alpha$ will produce a scatter of 5.2 orders of magnitude in $R_{\rm UV}$ (cf. Eq. \ref{eq:ruv1}), which is in rough agreement with the $\sim$ 6-order-of-magnitude scatter in our sample, cf. the top panel of Fig.\ \ref{llagnsample}.

Finally we examine the $R_{\rm UV}$--$\alpha_{\rm ox}$ relationship. With substitution of $\lambda$ in Eqs. \ref{eq:ruv1} \& \ref{eq:alphaox1}, one may quickly find a positive correlation between $R_{\rm UV}$ and $\alpha_{\rm ox}$. We note that this is actually misleading, as it only reflects the dependence on bolometric luminosity $\lbol$. Considering the large scatters in both $R_{\rm UV}$ and $\alpha_{\rm ox}$, the impact of $\lambda$ on the $R_{\rm UV}$ versus $\alpha_{\rm ox}$ is insignificant, e.g., the different symbols in Fig.\ \ref{r-ox} overlaps with one another. This actually inspires us to consider other factors, where we attribute to the viscosity parameter $\alpha$.
Since $R_{\rm UV}$ has a negative correlation with $\alpha$, while $\alpha_{\rm ox}$ has a positive relationship with $\alpha$ (cf. Eqs.\ \ref{eq:ruv1} \& \ref{eq:alphaox1}), we expect to have a negative correlation between $R_{\rm UV}$ and $\alpha_{\rm ox}$. Eliminating $\alpha$ in Eqs.\ \ref{eq:ruv1} \& \ref{eq:alphaox1}, we have,
\begin{equation}
\log R_{\rm UV} \approx -3.25\ \alpha_{\rm ox} - 0.95\ \log\lambda + const., \label{eq:3par}
\end{equation}
where $b\approx 2$ is adopted. Obviously, a strong negative $R_{\rm UV}$--$\alpha_{\rm ox}$ correlation is reproduced. The slope agrees with the observed value ($-2.39 \pm 0.48$, cf. Eq.\ \ref{eq:ruvobs}) at $2\sigma$ level. However, we admit that the moderate strong dependence on $\lambda$ predicted by this simple theoretical model disagrees with observations.

\section{SUMMARY}\label{summary}

In this work, we gather from literature a sample of LLAGNs whose X-rays are likely from the emission of hot accretion flows. The luminosity range of this sample is $10^{-6}\la\lambda\la 10^{-3}$, i.e. we exclude from our sample sources whose X-rays are likely dominated by emission from jet, and also moderately bright LLAGNs whose UV emission may have contributions from outer SSD. Our LLAGN sample includes 32 sources, among which 14 sources are low-$\lambda$ Seyfert galaxies and the rest 18 sources are LINERs. From this sample, we observe a strong negative correlation between $R_{\rm UV}$ and $\alpha_{\rm ox}$ (cf. Eq.\ \ref{eq:ruvobs}), and this correlation is insensitive to the Eddington ratio $\lambda$.

Then, based on the truncated accretion--jet model, we provide comprehensive understanding to this new observation, together with the other two negative correlations reported in literature, i.e. $R_{\rm X}$--$\lambda$ and $\alpha_{\rm ox}$--$\lambda$ correlations (note that we produce a null $R_{\rm UV}$--$\lambda$ relationship with large uncertainty). The scatter in these correlations derived is argued to relate to the viscosity parameter $\alpha$ of the hot accretion flow component.

From theoretical point of view, viscosity parameter determines the critical luminosity of hot accretion flow, above which it will enter into the ``physically-bright AGN'' accretion mode \citep{x2012,q2013,y2014}. More importantly, it is argued based on MHD simulations that, $\alpha$ has a strong dependence on the net vertical magnetic flux and/or magnetic field configuration carried by the accreted material (e.g., \citealt{h1995, b2008, bs2013, s2016}). For example, the value of $\alpha$ can vary about two order of magnitudes ($\sim0.01 - 1$) when the magnetic field strength $\beta$ (the ratio of gas to magnetic pressure) varies from $10^5$ to $10$ \citep{bs2013, s2016}. In this sense, deeper understanding on the correlation as well as the scatter of the $R_{\rm UV}$--$\alpha_{\rm ox}$ relationship will hopefully reveal the strength and/or configuration of magnetic field in hot accretion flows, which will also help to constrain the launching mechanism of jet (see e.g., \citealt{b1982, h2015}). Detailed numerical calculations are devoted to our future work.

\section*{ACKNOWLEDGEMENTS}
We appreciate the referee for valuable comments, and Minfeng Gu for useful discussions. We are supported in part by the National Program on Key Research and Development Project of China (Grant Nos. 2016YFA0400804 and 2016YFA0400704), the Youth Innovation Promotion Association of Chinese Academy of Sciences (CAS) (ids. 2015216 and 2016243), the Natural
Science Foundation of China (grants 11373056, 11233006, 11573051, 11633006 and 11661161012), the Natural Science Foundation of Shanghai (grant 17ZR1435800), the Key Research Program of Frontier Sciences of CAS (No. QYZDJ-SSW-SYS008). This work has made extensive use of the NASA/IPAC Extragalactic Database (NED), which is operated by the Jet Propulsion Laboratory, California Institute of Technology, under contract with the National Aeronautics and Space Administration (NASA).



\end{document}